\documentclass{article}
\usepackage{spconf,amsmath,amssymb,graphicx}
\usepackage{bm,here,cite}

\newcommand{\mymin}{\mathop{\rm min}\limits}

\def\rnum#1{\expandafter{\romannumeral #1}} 
\def\Rnum#1{\uppercase\expandafter{\romannumeral #1}} 

\def\@citex[#1]#2{%
  \let\@citea\@empty
  \@cite{\@for\@citeb:=#2\do
    {\@citea\def\@citea{], [}%
     \edef\@citeb{\expandafter\@firstofone\@citeb\@empty}%
     \if@filesw\immediate\write\@auxout{\string\citation{\@citeb}}\fi
     \@ifundefined{b@\@citeb}{\mbox{\reset@font\bfseries ?}%
       \G@refundefinedtrue
       \@latex@warning
         {Citation `\@citeb' on page \thepage \space undefined}}%
       {\hbox{\csname b@\@citeb\endcsname}}}}{#1}}

\title{MODE DOMAIN SPATIAL ACTIVE NOISE CONTROL USING SPARSE SIGNAL REPRESENTATION}
\name{Yu Maeno$^{\star}$, Yuki Mitsufuji$^{\star}$, and Thushara D. Abhayapala$^{\dagger}$}
\address{$^{\star}$Sony Corporation, Minato-ku, Tokyo, Japan \\
		$^{\dagger}$Research School of Engineering, The Australian National University, Canberra ACT 2601, Australia}

\begin{document}
\ninept
\maketitle
\begin{abstract}
Active noise control (ANC) over a sizeable space requires a large number of reference and error microphones to satisfy the spatial Nyquist sampling criterion, which limits the feasibility of practical realization of such systems. This paper proposes a mode-domain feedforward ANC method to attenuate the noise field over a large space while reducing the number of microphones required.
We adopt a sparse reference signal representation to precisely calculate the reference mode coefficients. The proposed system consists of circular reference and error microphone arrays, which capture the reference noise signal and residual error signal, respectively,
and a circular loudspeaker array to drive the anti-noise signal.
Experimental results indicate that above the spatial Nyquist frequency,
our proposed method can perform well compared to a conventional methods.
Moreover, the proposed method can even reduce the number of reference microphones while achieving better noise attenuation.
\end{abstract}
\section{Introduction}
\label{sec:intro}
Spatial ANC systems aim to attenuate undesired noise over a spatial region by generating an anti-noise sound field over the region using secondary sources.  Such a system is viewed as an extension to the well-studied single-channel ANC~\cite{kuo1995active,kuo1999active,kajikawa2012recent} to a multi-channel system~\cite{kuo1995active,elliott1987multiple}. However, to control a sound field over an extended region needs a large number of reference and error microphones, to capture the reference noise signals and residual error signals, respectively, as well as many loudspeakers to generate the anti-noise.  The requirement to have many microphones limits the viability of using spatial ANC systems in practice. 
Nevertheless, in most practical
applications~\cite{chen2015spatial,samarasinghe2016recent},
the underlying noise field is due to a small number of underlying noise sources.  In this paper, we aim to reduce the number of microphones required by spatial ANC systems by using a sparse signal representation.

%
 In practice, multiple-channel ANC in the frequency domain~\cite{douglas1999fast,bouchard2003computational} is widely used to
attenuate the noise field at multiple points where the error microphones are placed. One drawback of this approach to create a large quite zone is the requirement to uniformly place many error microphones inside the control region. In contrast, spatial-sound-field representation techniques, such as wave field synthesis (WFS)~\cite{berkhout1993acoustic} and higher order ambisonics (HOA)~\cite{poletti2000unified,abhayapala2002theory} are promising techniques to control not only multiple points but the entire space of interest. These techniques have been applied to ANC and have shown that mode-domain ANC can achieve noise attenuation over a large space even with less computational complexity~\cite{morgan1991adaptive,spors2008efficient,zhang2015noise}.
However,  such systems still require a large number of microphones and loudspeakers to reproduce the sound field properly, thus limiting the feasibility of practical implementations.  Without a theoretically sufficient number of sensors, the system suffers from artifacts due to spatial aliasing~\cite{ahrens2012analytic} due to violating the spatial Nyquist sampling criterion.

One way to alleviate the aliasing restriction is by expressing the noise field using only a small  number of basis functions from an over complete dictionary.
This approach is well known as compressive sensing (CS)~\cite{candes2008introduction}, which can provide an accurate solution for underdetermined problems where the sparse characteristic of the underlying signal field is usually used to solve the problem~\cite{malioutov2005sparse}.

In this paper, we propose a mode-domain ANC system that can perform noise attenuation in large space beyond the spatial Nyquist frequency using few reference microphones than required by Nyquist theory.  We represent the reference noise field as a superposition of weighted plane waves impinging from the far field. Using the estimated plane wave weights calculated by adopting several types of CS methods, we reconstruct the reference mode coefficients,
which are expected to include less artifacts caused by spatial aliasing. The proposed methods are evaluated and compared in terms of reference-noise-field reproduction accuracy and noise attenuation level. We demonstrate that our proposed method can overcome the spatial Nyquist frequency limitation. Furthermore, we show that the number of reference microphones can be reduced while controlling the noise field properly.

Nicolas~\cite{epain2009application} proposed CS sound field reproduction based on plane wave decomposition by introducing a HOA-based constraint.
However, they calculated the loudspeaker weights using amplitude-panning, which cannot be used directly for mode-domain processing.
Jihui~\cite{zhangg2016sparse,zhang2016multichannel} proposed the sparse complex FxLMS algorithm although they applied the CS approach to calculate sparse loudspeaker weights so that the spatial-aliasing artifacts cannot be avoided, which is caused by the spatial sampling of the microphone array.


%
\section{Problem statement}
\label{sec:formula}
\begin{figure}[t]
\vspace*{-5pt}
  \centering
  \centerline{\includegraphics[width=0.7\hsize]{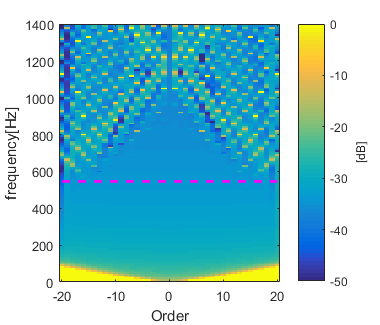}}
\vspace{-1pt}
\caption{Extracted circular mode coefficients of cylindrical wave using microphone outputs. The horizontal broken line indicates the spatial Nyquist frequency.}
\label{fig:spectrogram}
\vspace{-4pt}
\end{figure}

Let us consider a circular microphone array capturing a noise field generated by noise sources outside the microphone array in 2D space.
We can represent an incident noise field at an arbitrary
point ${\bf x} \equiv (r, \phi)$:
\begin{eqnarray}
S({\bf x},k) &=& \sum^{\infty}_{m=-\infty}\beta_m(k) J_m(kr) e^{im \phi} \label{eq:cylex1} \\
&\approx& \sum^{M}_{m=-M}\beta_m(k) J_m(kr) e^{im \phi} ,
\label{eq:cylex2}
\end{eqnarray}
where $k = 2 \pi f / c$ is the wave number, $f$ is the frequency, $c$ is the speed of sound, $\beta_m(k)$ is the $m$th-order circular mode coefficient, and $J_m(\cdot)$ is the $m$th-order Bessel function. In practice, (\ref{eq:cylex1}) can be truncated to $M = \lceil ekR/2 \rceil$~\cite{kennedy2007intrinsic} or $M = \lceil kR \rceil$~\cite{ward2001reproduction}, where $R$ is the radius of a controllable spatial region.
The mode coefficient can be calculated utilizing orthogonal exponential functions:
\begin{equation}
\beta_m(k) = \frac{1}{Q J_m(kr)} \sum^{Q}_{q=1} S({\bf x},k) e^{-im \phi_q} ,
\label{eq:mode_extract}
\end{equation}
where $Q$ is the number of microphones and $\phi_q$ is the azimuth angle
of the $q$th microphone.
We need at least $Q \ge 2M+1$ microphones to avoid spatial aliasing~\cite{samarasinghe2014wavefield}.
Figure~\ref{fig:spectrogram} shows the mode coefficients of
the noise field captured by a  $41$-element circular microphone array of radius $2$~m.
The maximum order of the mode coefficients is $M=20$.
Observe that spatial aliasing artifacts appear
for higher modes above 546 Hz, which is the spatial Nyquist frequency for this condition.
The high amplitude of low frequency corresponds to 
an evanescent wave~\cite{williams1999fourier} of the cylindrical source.
This evanescent component only exists near the source, and the amplitude decays exponentially~\cite{ahrens2012analytic}.
Therefore, we assume that
the impact on the performance of an ANC system inside the control region
is minimal. 

Our objective is to derive an algorithm to acquire precise reference mode coefficients of a noise field, which are then used to perform ANC
instead of using direct mode coefficient extraction (\ref{eq:mode_extract}) containing
spatial aliasing artifacts caused by the theoretical restriction, i.e. $M = \lceil kR \rceil$.
We expect the noise field can be attenuated at higher frequencies beyond the spatial Nyquist frequency, even with an insufficient number of microphones.

\begin{figure}[t]
\vspace*{-5pt}
  \centering
  \centerline{\includegraphics[width=0.95\hsize]{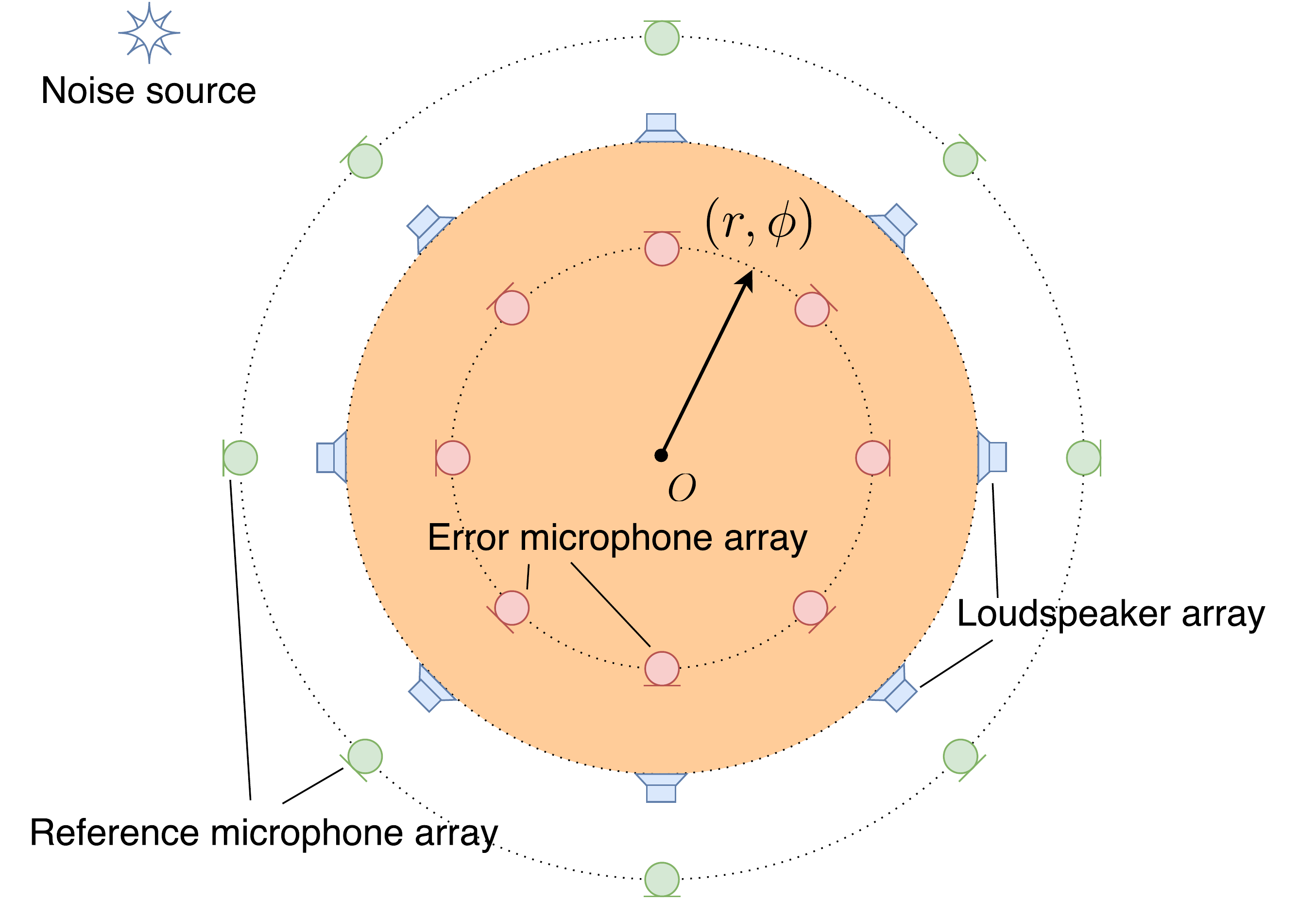}}
\vspace{-1pt}
\caption{Geometric setup of feedforward ANC system. The colored circle indicates the control region.}
\label{fig:NCsys}
\vspace{-4pt}
\end{figure}
%

%
\section{Plane wave weight estimation}
\label{sec:estimation}
In this section, we derive mode coefficient reconstruction based on plane wave decomposition
assuming that the noise field is constructed by a superposition of weighted plane waves.
The weights of the plane waves can be estimated accurately by utilizing
the CS approach, which can be used 
to derive a sparse solution using only a few measurements.

As an alternative to (\ref{eq:cylex2}), the noise field can be described by
$L$ plane waves impinging towards the coordinate origin~\cite{fan2014practical}:
\begin{equation}
S({\bf x}, k) = \sum^{L}_{\ell=1} \gamma_\ell e^{ik \hat{ {\bf y}}_\ell \cdot {\bf x}} ,
\label{eq:pwdecomp}
\end{equation}
where $\gamma_\ell$ is the weight of $\ell$th plane wave and $\hat{ {\bf y}}_\ell \equiv (1, \phi_\ell)$ is a
unit vector towards the direction of the $\ell$th plane wave.
Circular expansion of the incident noise field due to an unit-magnitude plane wave is given by
\begin{equation}
e^{ik \hat{ {\bf y}}_\ell \cdot {\bf x}} = \sum^M_{m=-M} i^m e^{-im \phi_\ell} J_m(kr) e^{im \phi} .
\label{eq:pw}
\end{equation}
From (\ref{eq:pwdecomp}) and (\ref{eq:pw}), we have
\begin{equation}
S({\bf x}, k) = \sum^M_{m=-M} \underbrace{ i^m \sum^{L}_{\ell=1} \gamma_\ell e^{-im \phi_\ell} }_{\beta_m(k)} J_m(kr) e^{im \phi} .
\label{eq:coefreconst}
\end{equation}
Equation~(\ref{eq:coefreconst}) shows that the mode coefficient $\beta_m(k)$ can be reconstructed using the plane wave weight $\gamma_\ell$
to reproduce the incident noise field.

In order to apply CS and estimate a sparse solution
in (\ref{eq:pwdecomp}),
we represent it in matrix form:
\begin{equation}
\bm{s} = \bm{E} \bm{\gamma} ,
\label{eq:matrix}
\end{equation}
where $\bm{s}$ is a $Q \times 1$ vector of the observed signals, $\bm{\gamma}$ is an $L \times 1$ vector of plane wave weights, and $\bm{E}$ is a
$Q \times L$ matrix given by
\begin{equation}
\bm{E} = 
\begin{bmatrix}
e^{ik \hat{ {\bf y}}_1 \cdot {\bf x}_1} & \cdots & e^{ik \hat{ {\bf y}}_L \cdot {\bf x}_1} \\
\vdots & \ddots & \vdots \\
e^{ik \hat{ { \bf y}}_1 \cdot {\bf x}_Q} & \cdots & e^{ik \hat{ {\bf y}}_L \cdot {\bf x}_Q}
\end{bmatrix}
.
\label{eq:matE}
\end{equation}
%
We assume that the composition of plane waves is sparse with the underdetermined condition $(Q<L)$ in (\ref{eq:matrix}).

\begin{figure}[t]
\vspace*{-10pt}
  \centering
  \centerline{\includegraphics[width=8.5cm]{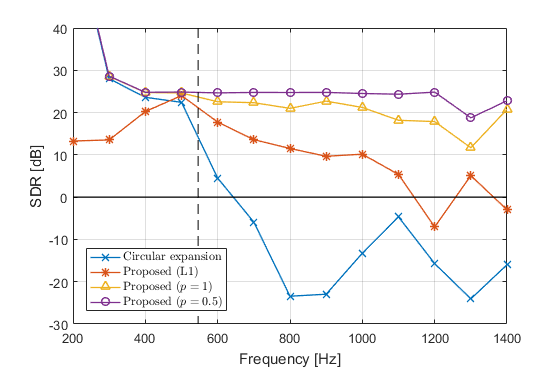}}
\vspace{-1pt}
\caption{SDR of reference sound field at several frequency bins. The vertical broken line indicates the spatial Nyquist frequency corresponding to the reference microphone array.}
\label{fig:SDR}
\vspace{-4pt}
\end{figure}

\subsection{$\ell_1$-norm constrained minimization}

One of the most widely used technique to estimate a sparse solution
is to minimize a cost function with an $\ell_1$ constraint
on the plane wave weights:
\begin{equation}
\mathcal{J} = \| \bm{v} \|^2_2 + \lambda_1 \| \bm{\gamma} \|_1 ,
\label{eq:l1const}
\end{equation}
where $\bm{v} = \bm{s} - \bm{E} \bm{\gamma}$ and $\lambda_1$ controls the strength
of the sparsity constraint.
By utilizing the steepest descent algorithm,
we can update the weights $\bm{\gamma}$ as 
\begin{equation}
\bm{\gamma}(n+1) = \bm{\gamma}(n) - \frac{\mu_{\rm mic}}{2} \nabla \mathcal{J}(n) ,
\label{eq:l1update}
\end{equation}
where $n$ is the iteration index and $\mu_{\rm mic}$ is the step size.
A gradient of the cost function $\mathcal{J}(n)$ can be calculated as~\cite{zhangg2016sparse,zhang2016multichannel}
\begin{equation}
\nabla \mathcal{J}(n) = 2 \bm{E}^H \bm{v}(n)
 + \lambda_1 \Bigl( {\rm sgn} \bigl\{ \mathfrak{R}\left[\bm{\gamma}(n)\right] \bigr\}
 + i {\rm sgn} \bigl\{ \mathfrak{I} \left[(\bm{\gamma}(n)\right] \bigr\} \Bigr) ,
\label{eq:l1deriv}
\end{equation}
where $(\cdot)^H$ denotes the Hermitian-transpose operation, ${\rm sgn(\cdot)}$ denotes the sign function,
and $\mathfrak{R}(\cdot)$ and $\mathfrak{I}(\cdot)$ denote the real and imaginary parts of the argument.
Finally, substituting (\ref{eq:l1deriv}) into (\ref{eq:l1update}), the plane-wave weights can be calculated as:
\begin{multline}
\bm{\gamma}(n+1) = \bm{\gamma}(n) - \mu_{\rm mic} \bm{E}^H \bm{v} (n) \\
 - \frac{1}{2} \mu_{\rm mic} \lambda_1 \Bigl( {\rm sgn} \bigl\{ \mathfrak{R}\left[\bm{\gamma}(n)\right] \bigr\}
 + i {\rm sgn} \bigl\{ \mathfrak{I} \left[(\bm{\gamma}(n)\right] \bigr\} \Bigr) .
\label{eq:l1update2}
\end{multline}

\subsection{Iteratively reweighted least squares (IRLS)}

It is known that by introducing the $\ell_p$-norm and solving
the following optimization problem, where $0<p<1$
can provide a sparse solution with much fewer measurements than $p=1$
~\cite{chartrand2007exact,chartrand2008iteratively}:
\begin{equation}
\mymin_{\bm{\gamma}} \| \bm{\gamma} \|_p^p, \mbox{ s.t. } \bm{s} = \bm{E} \bm{\gamma} .
\label{eq:lpconst}
\end{equation}
Applying basis pursuit denoising (BPD)~\cite{carrillo2009iteratively}, (\ref{eq:lpconst}) can be written in
an unconstrained form as:
\begin{equation}
\mymin_{\bm{\gamma}} \frac{1}{p} \| \bm{\gamma} \|^p_p + \frac{\lambda_2}{2} \| \bm{v} \|^2_2 .
\label{eq:lpunconst}
\end{equation}
Although the optimization problem (\ref{eq:lpunconst}) is a nonconvex function,
we can adopt IRLS~\cite{chartrand2008iteratively,carrillo2009iteratively}
to solve it iteratively.
IRLS is known to converge relatively fast in terms of the number of iterations,
however, the algorithm includes matrix inversion in every iteration,
which may lead to high computational cost.
There is a fast implementation of the IRLS algorithm~\cite{chen2014fast}
to deal with this issue.

%
\section{Mode-Domain Active Noise Control}
\label{sec:ANC}

\begin{figure}[t]
\vspace*{-10pt}
  \centering
  \centerline{\includegraphics[width=1.0\hsize]{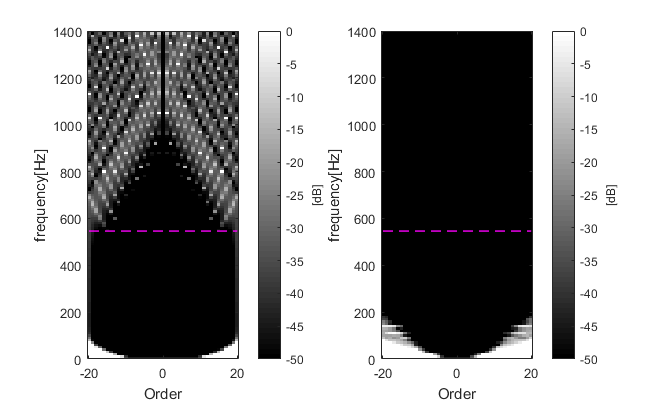}}
\begin{minipage}{0.49\linewidth}
\centering
(a)
\end{minipage}
\begin{minipage}{0.49\linewidth}
\centering
\hspace{-11mm}
(b)
\end{minipage}
\vspace{-1pt}
\caption{Mode coefficients error between original and reconstructed noise field. The horizontal broken line indicates the spatial Nyquist frequency. (a)~MDFF. (b)~Proposed $(p=0.5)$.}
\label{fig:spec_err}
\vspace{-4pt}
\end{figure}

In this section, we describe a mode-domain adaptive filtering algorithm
based on the filtered-x least mean square (FxLMS) algorithm~\cite{burgess1981active}.
Consider two circular microphone arrays placed in a free field, as depicted in Fig.~\ref{fig:NCsys}.
The outer circle is a reference microphone array capturing reference signals,
and the inner circle is an error microphone array, measuring the residual signals.
Between those two microphone arrays, a circular loudspeaker array is placed to
drive the anti-noise signals.

Adopting the mode-domain feedforward (MDFF) FxLMS algorithm,
a residual mode coefficient at step $n$ can be written as:
\begin{equation}
e_m(n, k) = \beta_m(k) + w_m(n, k) x'_m(n, k) ,
\label{eq:FxLMS}
\end{equation}
where $w_m(n, k)$ is the $m$th-order mode weight of the adaptive filter, $x'_m(n, k) = g_m(k) x_m(n, k)$ is the
$m$th-order filtered reference mode coefficient, $g_m(k)$ is the $m$th-order mode coefficient of the acoustic transfer function
of the loudspeaker, and $x_m(n, k)$ is the $m$th-order reference mode coefficient.
In the 2D free-field space, $g_m(k) = -\frac{i}{4} H_m^{(2)}(kR_s)$, where $H_m^{(2)}(\cdot)$ is the second-kind Hankel function of order $m$, and $R_s$ is the radius of the loudspeaker array. 
The reference mode coefficient can be calculated as
\begin{equation}
x_m(n, k) = i^m \sum^{L}_{\ell=1} \gamma_\ell(n, k) e^{-im \phi_\ell} ,
\label{eq:ref}
\end{equation}
using $\gamma_\ell(n, k)$ reconstructed from the reference signals by solving the optimization
problem mentioned in the previous section.
Finally, filtered-x normalized least mean square (FxNLMS)
~\cite{slock1993convergence} algorithm updates the mode weight as:
\begin{equation}
w_m(n+1, k) = w_m(n, k) - \mu_{\rm sp} \frac{e_m(n, k) \overline{x'_m(n, k)}}{\overline{x'_m(n, k)} x'_m(n, k)} ,
\label{eq:update}
\end{equation}
where $\overline{(\cdot)}$ denotes the complex conjugate operation and $\mu_{\rm sp}$ is the step size.

%
\section{Experiment}
\label{sec:exp}

\begin{figure}[t]
\vspace*{-10pt}
  \centering
  \centerline{\includegraphics[width=8.5cm]{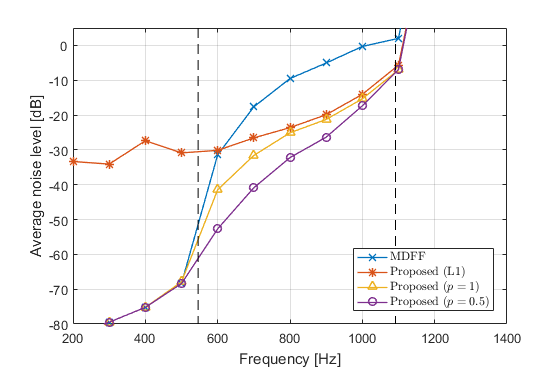}}
\caption{Average noise level at several frequency bins after 50 iterations. The vertical broken lines indicate the spatial Nyquist frequencies corresponding to the reference (left) and the error (right) microphone array.}
\label{fig:NoiseLevel}
\vspace{-3pt}
\end{figure}
%

%
%

We evaluated and compared the accuracy
of the reconstructed reference mode coefficients
and noise attenuation level among following methods:
(\rnum{1})~MDFF corresponds to the conventional method based on~(\ref{eq:mode_extract})
and~(\ref{eq:update}) as a baseline;
(\rnum{2})~Proposed~(L1) corresponding to the $\ell_1$-norm constrained ANC method
based on~(\ref{eq:l1update2});
(\rnum{3})~Proposed~$(p=1)$; and (\rnum{4})~Proposed~$(p=0.5)$, corresponding to the IRLS method
based on~(\ref{eq:lpunconst}).
Since reference microphones for ANC are typically placed outermost,
the reference-mode coefficients are more likely to be affected by spatial aliasing artefacts.
To overcome this problem, we applied the proposed method to the reference microphone array outputs.

The radii of the error and reference microphone arrays are $1$~m and $2$~m, respectively. Each microphone array consists of $41$ omni-directional microphones. The loudspeaker array consists of $41$ monopole loudspeakers and the radius is $1.5$~m.
The maximum order of the mode coefficients is $M=20$. For these experimental conditions, the spatial Nyquist frequency corresponding to
the reference microphone array is $546$~Hz, and to the error microphone array is $1092$~Hz.
We use $L=128$ plane waves to construct the over-complete matrix $\bm{E}$.

\subsection{Reference mode coefficient accuracy}
We calculated the signal-to-distortion ratio (SDR)~\cite{koyama2014sparse}
inside the control region depicted as a colored region
in Fig.~\ref{fig:NCsys} to measure the accuracy of
the noise field reproduced from the reference signals,
which is important to attenuate the noise field properly in feedforward ANC.
%
%

Figure~\ref{fig:SDR} shows the frequency dependent SDR calculated using several methods.
The circular expansion is based on (\ref{eq:mode_extract}).
The SDRs using the proposed method were calculated after convergence.
We can clearly see that above the spatial Nyquist frequency,
the circular expansion method can no longer reproduce the  noise field accurately,
while the proposed methods can provide better accuracy at higher frequencies.
Figure~\ref{fig:spec_err} depicts the mode coefficient error vs frequency and shows spatial aliasing artifacts.
There are higher errors appearing from higher modes above the spatial Nyquist frequency in MDFF.
Note that the higher errors in the evanescent region have only minimal impact on the ANC system, as previously mentioned.

Our preliminary experiment shows that the convergence speed vary among each method.
For example, Proposed~(L1) needs around 100 iterations to converge, whereas, Proposed~$(p=0.5)$ needs only around 10 iterations.

\subsection{Noise attenuation level}
We performed adaptive ANC by adopting several methods and
evaluating the noise attenuation level.
%
Figure~\ref{fig:NoiseLevel} shows the noise attenuation level at several frequencies after 50 iterations.
The vertical broken lines indicate the spatial Nyquist frequencies correspond to
the reference microphone array and the error microphone array, respectively.
Below the Nyquist frequency of the reference microphone array,
the MDFF and the proposed method gave almost the same results
except Proposed~(L1), which gave even worse performance.
We found that this was due to the strength of the constraint
that was controlled by the parameter $\lambda_1$.
The $\ell_1$ constraint was too strong to calculate accurate reference mode coefficients
for lower frequencies below the spatial Nyquist frequency.
We confirmed that the result improved once we manually tuned the parameter
for each frequency, however,
we use a fixed parameter in this paper since the issue is out of the scope
and also we are not tuning any parameters for other methods.

For higher frequencies, the attenuation level of MDFF degrades due to spatial aliasing,
however, the proposed method can still attenuate the noise.
Above the Nyquist frequency of the error microphone array,
all methods were out of control and no attenuation could be achieved.

\begin{figure}[t]
\vspace*{-10pt}
  \centering
  \centerline{\includegraphics[width=1.1\hsize]{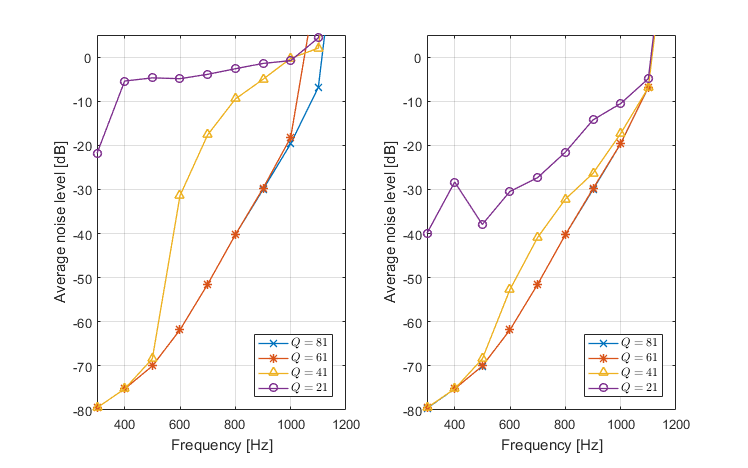}}
\begin{minipage}{0.49\linewidth}
\centering
(a)
\end{minipage}
\begin{minipage}{0.49\linewidth}
\centering
(b)
\end{minipage}
\caption{Average noise level among various setups for the number of reference microphones. (a)~MDFF. (b)~Proposed $(p=0.5)$.}
\label{fig:micnum}
\vspace{-3pt}
\end{figure}
%

%
%

\subsection{Reduction of the number of reference microphones}

When the noise field can be expressed by a sparse set of plane wave weights,
we can reduce the number of reference microphones to calculate the reference mode coefficients.
Figure~\ref{fig:micnum} shows the average noise level after 50 iterations of ANC.
Recalling the Nyquist frequency of the reference microphone array,
it is natural that the noise attenuation performance starts degrading
above $546$~Hz for $Q=41$ and $1092$~Hz for $Q=81$ in MDFF.
On the other hand, we can see that the proposed method performs well over frequencies
even though we reduced the number of reference microphones.
For both methods, the ANC system starts diverging above $1092$~Hz, which is the Nyquist frequency of the error microphone array.

\section{Conclusion}
\label{sec:conclusion}
In this paper, we explored the mode-domain feedforward ANC system performance using sparse signal representation for the reference signals.
We compared several CS methods and showed that the IRLS algorithm with $p=0.5$ performs well in terms of both convergence speed and noise attenuation level.
We showed that the proposed method can attenuate the noise field beyond the spatial Nyquist frequency even with fewer reference microphones than required by spatial sampling theory.

%
%
%
%


\bibliographystyle{IEEEbib}
\bibliography{ref}

\end{document}